\documentclass{PoS}

\usepackage[intlimits]{amsmath}
\usepackage{amssymb}
\usepackage{bbm}

\newcommand\T{\rule{0pt}{2.6ex}}
\newcommand\B{\rule[-1.2ex]{0pt}{0pt}}

\title{Meson spectroscopy with derivative quark sources}

\ShortTitle{Meson spectroscopy with derivative quark sources}

\author{Christof Gattringer$^a$, Leonid Ya. Glozman$^a$, C. B. Lang$^a$,
        \speaker{Daniel Mohler}\hspace{1mm}$^a$ \newline and Sasa Prelovsek$^b$\\
        \llap{$^a$} Institut f\"ur Physik, FB Theoretische Physik,
        Universit\"at Graz, A-8010 Graz, Austria\\
	\llap{$^b$} University of Ljubljana and Institute Josef Stefan, 1000 Ljubljana, Slovenia\\
        E-mail: \email{christof.gattringer@uni-graz.at},
        \email{leonid.glozman@uni-graz.at}, \email{christian.lang@uni-graz.at},
        \email{daniel.mohler@uni-graz.at}, \email{Sasa.Prelovsek@ijs.si}}

\abstract{We present results for masses of light mesons obtained
  with the variational method using an enhanced basis of interpolating field
  operators with different quark smearings. The interpolators are constructed
  from Jacobi-smeared quarks of a Gaussian type as well as from derivative
  quark sources obtained by a covariant derivative acting on the
  Gaussian sources. For our analysis we use quenched gauge configurations with
  Chirally Improved quarks and the L\"uscher-Weisz gauge action on a
  $16^3\times 32$ lattice with $a=0.148\,\mathrm{fm}$. We discuss the influence
  of derivative sources on the overlap with excited states. }

\FullConference{The XXV International Symposium on Lattice Field Theory\\
		 July 30 - August 4 2007\\
		 Regensburg, Germany}

\begin{document}

\section{Introduction}
Lattice QCD opened the possibility of providing a model independent {\it ab initio}
calculation of the QCD mass spectrum.
Extracting signals from excited states has however proven to be a formidable task. A previous study of the BGR-Collaboration \cite{Burch:2006dg} employed the variational
technique \cite{Michael:1985ne,Luscher:1990ck} to study excited mesons
on the lattice using standard meson interpolators with Gaussian-smeared sources
and sinks.\\
For an accurate representation of excited states in the variational approach it is crucial to use a
basis with good overlap with the physical states. We therefore construct additional
meson-interpolators with derivative sources obtained by a covariant
derivative acting on the Jacobi smeared sources and explore their effect in
the variational approach.

\section{Details of the calculation}

\subsection{The variational method}
We use a basis of interpolators $O_i$, $i=1,\dots,N$ with the quantum numbers
of the desired states (projected to zero momentum) and compute the matrix of
cross correlations:
\begin{align}
C(t)_{ij}&=\left<O_i(t)\overline{O_j}(0)\right>=\sum_n\bigl<0|O_i|n\bigr>\bigl
<n\bigl\lvert O_j^\dagger \bigr\rvert0\bigr >e^{-tM_n}\ .
\end{align}
It can be shown \cite{Luscher:1990ck} that the solutions to the generalized
eigenvalue problem
\begin{align}
C(t)\vec{v}_i&=\lambda_i(t)C(t_0)\vec{v}_i\ ,
\end{align}
behave as
\begin{align}
\lambda_i(t)\propto e^{-tM_i}\left(1+\mathcal{O}\left(e^{-t\Delta M_i}\right)\right)\ ,
\end{align}
where $\Delta M_i$ is the mass difference between the state $i$ and the
closest lying state.\\
Therefore, up to the given order, the largest eigenvalue decays with the mass
of the ground state, the second largest eigenvalue with the mass of the first
excited state and so on.
For an accurate description, the interpolators should be linearly independent,
as orthogonal as possible and possess a strong overlap with physical
states. The task will be to construct new interpolators which are independent
of the standard interpolators and have a good overlap with excited states.

\subsection{Effective masses and eigenvectors}
To display the results, one commonly plots the \emph{effective masses}
determined from ratios of eigenvalues
\begin{align}
a\,M_{i,\,\mathrm{eff}}\left(t+\frac{1}{2}\right) &=
\mathrm{ln}\left(\frac{\lambda_i(t)}{\lambda_i(t+1)}\right)\ .
\end{align}
A fit to the exponential decay of the eigenvalues is then performed in the time interval where the
effective mass plot shows a plateau. As a consistency check, the entries of
the corresponding eigenvectors should also show plateaus and can serve as a
fingerprint for the ``wave functions''. To deal with eigenvectors that are
orthonormal we calculate the eigenvectors of the modified eigenvector problem
\begin{align}
C(t_0)^{-\frac{1}{2}}C(t)C(t_0)^{-\frac{1}{2}}\vec{v}^{\,\prime}_i&=\lambda_i(t)\vec{v}^{\,\prime}_i\ .
\end{align}
(This assumes $C(t_0)$ to be positive definite.) 

\subsection{Smeared sources and sinks}
The first step for all our sources is Jacobi smearing
\cite{Gusken:1989ad, Best:1997qp} of point sources $s_0$ located at timeslice $t=0$:
\begin{align}
s_0^{(\alpha,a)}(\vec{y},0)_{\rho,c}&=\delta(\vec{y},0)\delta_{\rho
  \alpha}\delta_{ca}\ ,\\
s^{(\alpha,a)}&=\sum_{n=0}^N\kappa^nH^n\ ,\\
H(\vec{x},\vec{y})&=\sum_{i=1}^3\left(U_i(\vec{x},0)\delta(\vec{x}+\hat{i},\vec{y})+U_i(\vec{x}-\hat{i},0)^\dagger\delta(\vec{x}-\hat{i},\vec{y})\right)\ .
\end{align}
In \cite{Burch:2006dg} two different parameter values for $\kappa$ and $N$,
giving rise to wide and narrow sources ($S_w$,$\,S_n$), have been used. In addition, we now construct covariant derivatives which act upon a
wide smeared source
to form our derivative quark sources $W_{d_i}$:
\begin{align}
P_i(\vec{x},\vec{y})&=U_i(\vec{x},0)\delta(\vec{x}+\hat{i},\vec{y})-U_i(\vec{x}-\hat{i},0)^\dagger\delta(\vec{x}-\hat{i},\vec{y})\
,\\
W_{d_i}&=P_iS_w\ .
\end{align}
With these sources, meson interpolators of definite quantum numbers are constructed.

\subsection{Interpolators used}
Table 1 shows the interpolators used for different meson channels. The ones
labeled ``old interpolators'' are those of \cite{Burch:2006dg}, while the
``new interpolators'' contain at least one derivative. The total number of
interpolators as given in the last column can be obtained by combining all
non-degenerate combinations of the different smearings with all possible Dirac structures.

In some cases, an (anti-) symmetrization of the interpolators is necessary to obtain the
correct behavior under charge conjugation. Therefore, interpolators denoted as
$\bar{u}_{d_i}\Gamma d_{n/w}$ in the table should be read as $\bar{u}_{d_i}\Gamma d_{n/w}-\bar{u}_{n/w}\Gamma d_{d_i}$.
We restrict ourselves to light, isovector ($I=1$) mesons with degenerate
quark masses $m_u=m_d$.

\begin{table}[h!]
\begin{center}
\hspace*{-4mm}
\begin{tabular}{|c|c|c|c|c|c|c|c|}
\hline
 \T\B&$J^{PC}$&\multicolumn{2}{c|}{\emph{old
 interpolators}}&\multicolumn{3}{c|}{\emph{additional, new interpolators}}&$\sharp$\\
\hline
\T\B pseudoscalar&$0^{-+}$&$\bar{u}_{n/w}\gamma_5d_{n/w}$&$\bar{u}_{n/w}\gamma_4\gamma_5d_{n/w}$&$\bar{u}_{d_i}\gamma_i\gamma_4\gamma_5d_{n/w}$&$\bar{u}_{d_i}\gamma_5d_{d_i}$&$\bar{u}_{d_i}\gamma_4\gamma_5d_{d_i}$&10\\
\hline
\T\B scalar&$0^{++}$&$\bar{u}_{n/w}d_{n/w}$&&$\bar{u}_{d_i}\gamma_id_{n/w}$&$\bar{u}_{d_i}\gamma_i\gamma_4d_{n/w}$&$\bar{u}_{d_i}d_{d_i}$&8\\
\hline
\T\B vector&$1^{--}$&$\bar{u}_{n/w}\gamma_id_{n/w}$&$\bar{u}_n\gamma_i\gamma_4d_{n/w}$&$\bar{u}_{d_i}d_{n/w}$&$\bar{u}_{d_i}\gamma_kd_{d_i}$&&9\\
\hline
\T\B pseudovector&$1^{++}$&$\bar{u}_{n/w}\gamma_i\gamma_5d_{n/w}$&&$\bar{u}_{d_i}\gamma_k\gamma_5d_{d_i}$&&&4\\
\hline
\end{tabular}
\end{center}
\caption{Meson interpolators; $n/w$ denote narrow and wide Jacobi
smearing and $d_i$ stands for a derivative source in
the $i$-direction. The last column shows the total number of
different interpolators.}
\end{table}

\subsection{Technicalities}
For our analysis we used 99 uncorrelated quenched gauge
configurations with Chirally Improved (CI)
\cite{Gattringer:2000qu,Gattringer:2000js} quarks and the L\"uscher-Weisz
gauge action \cite{Luscher:1984xn}. We work on a $16^3\times 32$ lattice with
$a=0.148\ \mathrm{fm}$ determined from \cite{Gattringer:2001jf} the Sommer parameter. We study
several quark mass parameters in the range $am_q=0.02\dots 0.2$. All errors we
quote are statistical errors determined with the jackknife method.

\section{Results}
To get an idea about the properties of the correlators, we first plot the
normalized diagonal entries of the correlation matrices for the pseudoscalar
and vector mesons. In figure 1 the simple interpolators have been colored blue, while the new correlators from
derivative sources appear in red. As can be seen from the behavior at small
euclidean times $t$, the new
correlators have stronger contributions from excited states, i.e., they start
out with a steeper slope. However,
all correlators are dominated by the ground state at large time separations.

\begin{figure}[h!]
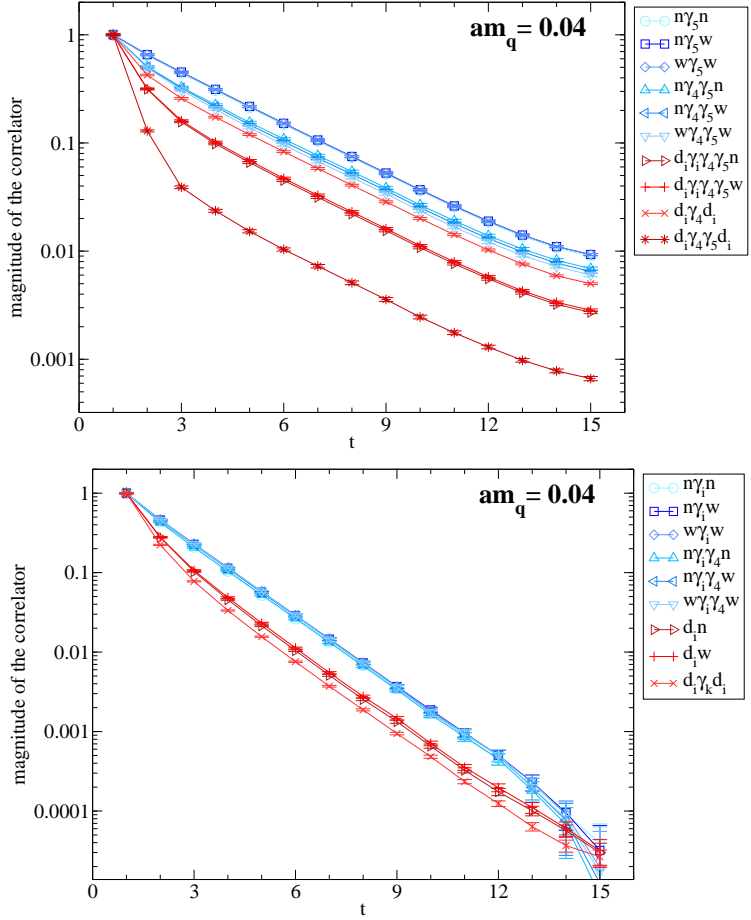

\begin{center}
\includegraphics[scale=0.37,clip]{pion_corr_normalized_0.04.eps}\\[0.2cm]
\includegraphics[scale=0.37,clip]{rho_corr_normalized_0.04.eps}
\end{center}
\vspace*{-2mm}
\caption{Diagonal correlators for the pseudoscalar (top) and vector
(bottom) mesons at $am_q=0.04$. $n/w$ denote narrow and wide Jacobi
smearing and $d_i$ stands for a derivative source in
the $i$-direction.}
\end{figure}

Figure 2 shows a comparison of selected effective mass plateaus from the optimal
set of old interpolators compared to the new results (obtained from the
optimal combination of both old and new interpolators) at different values of the quark
mass. Plateaus are also found in the components of the corresponding
eigenvectors.

For the pseudoscalar mesons (graphs on the upper l.h.s.) we obtain improved results for
the excited states at all quark masses. Both the old and the new sets
are obtained with a basis of four interpolators. This shows that the observed effect
does not just stem from simply enlarging the basis. Some interpolators mainly
contribute noise without significantly enhancing the overlap to the physical states.
Looking at the modified eigenvectors of equation (2.5) as depicted in figure 3, clear plateaus are observed and the new
interpolators (second and third entries) mainly contribute to the second excited state which could not be
observed using the old set of interpolators alone.

Similarily for the vector mesons (upper r.h.s.\ of figure 2), improved results are obtained
for the excited states at all quark masses. Here, however, the new basis
consists of six interpolators while the old basis only contains four. Both
the ground state and newly observed third excited state are dominated by the
old interpolators while the first and second excited states are dominated by
the new interpolators.We also would like to note that the finer ($a\approx 0.12\,\mathrm{fm}$) quenched
lattice considered in \cite{Burch:2006dg} allowed for a cleaner separation of
the first and second excited states.

For the scalar (lower l.h.s.\ of figure 2) and pseudovector (lower r.h.s.\ of figure 2) mesons, there is a noticeable improvement of the ground state
plateaus. The old graph for the pseudovectors stems from a basis of three
interpolators while the new graph is obtained from just one correlator
alone. The location of the plateau agrees within error bounds but the errors
obtained from the new interpolator are considerably smaller. For the scalars,
the plots have been obtained from a basis of three interpolators. Using
derivative sources, plateaus can be extracted at small values of the quark
mass where no plateau is observed in the old interpolators. For the quenched
data, scalar correlators contain contaminations from ghost states and it has
been shown, that the variational method can isolate such unphysical
contributions \cite{Burch:2005wd}. Overall, ghost contributions to derivative
interpolators seem to be smaller and some interpolators show almost no sign of
such contributions. Where present, the variational method separates
the ghosts from the physical signal and the ground state plateau can be clearly
extracted (shown in figure 2). Concerning a full dynamical simulation, data from only a few dynamical
configurations suggests that interpolators of the type
$\bar{u}_{d_i}\gamma_id_{n/w}$, which are important in the quenched case,
seem to have very limited overlap in the dynamical case.

\begin{figure}[t!]
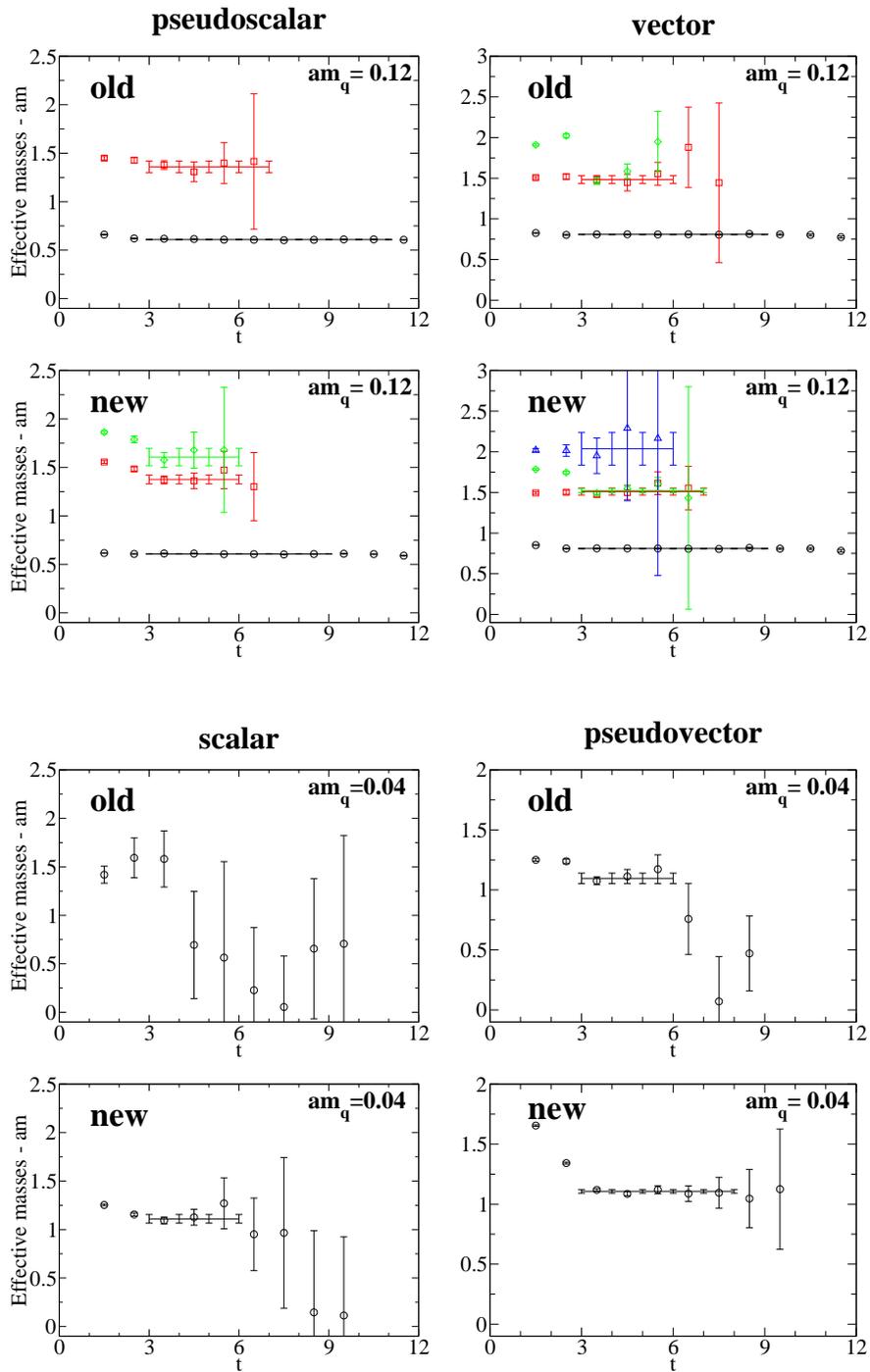

\begin{center}
%\includegraphics[scale=0.23]{masses_100111000000_m0.12.eps}
%\hspace{0.3cm}\includegraphics[scale=0.23]{masses_vector_100111000000_m0.12.eps}
%\hspace{0.3cm}\includegraphics[scale=0.23]{masses_axial_11100000000_m0.12.eps}\\[0.5cm]
%\includegraphics[scale=0.23]{masses_100001001100_m0.12.eps}
%\hspace{0.3cm}\includegraphics[scale=0.23]{masses_vector_11100011001_m0.12.eps}
%\hspace{0.3cm}\includegraphics[scale=0.23]{masses_axial_00000000001_m0.12.eps}\\
\includegraphics[scale=0.27,clip]{masses_100111000000_m0.12.eps}
\hspace{0.3cm}\includegraphics[scale=0.27,clip]{masses_vector_100111000000_m0.12.eps}\\[0.2cm]
\includegraphics[scale=0.27,clip]{masses_100001001100_m0.12.eps}
\hspace{0.3cm}\includegraphics[scale=0.27,clip]{masses_vector_11100011001_m0.12.eps}\\[0.8cm]
\includegraphics[scale=0.27,clip]{masses_scalar_111000000000_m0.04.eps}
\hspace{0.3cm}\includegraphics[scale=0.27,clip]{masses_axial_11100000000_m0.12.eps}\\[0.2cm]
\includegraphics[scale=0.27,clip]{masses_scalar_000000110010_m0.04.eps}
\hspace{0.3cm}\includegraphics[scale=0.27,clip]{masses_axial_00000000001_m0.12.eps}
\end{center}
\caption{Effective mass plots for the pseudoscalar (upper left), vector
(upper right), scalar (lower left) and pseudovector (lower right) mesons. We compare the results from the best selection of simple interpolators (``old'')
to the case where the new derivative interpolators are added (``new''). The
horizontal lines indicate the fit ranges and their error bars indicate the
statistical error of the fit.}
\end{figure}

\begin{figure}[h!]
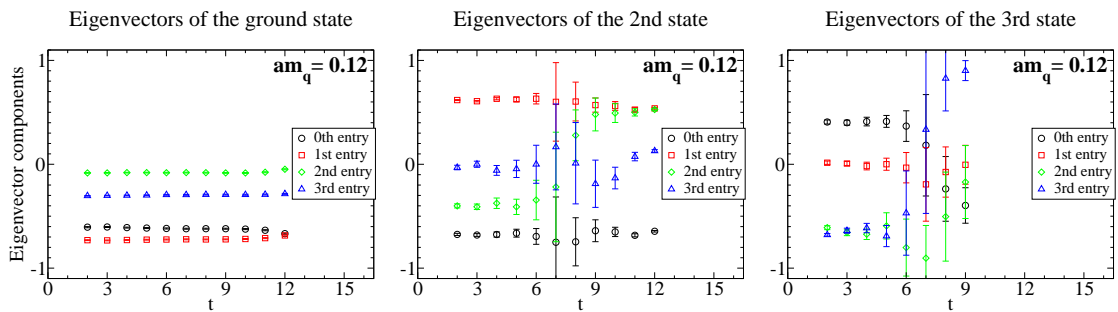

\begin{center}
\includegraphics[scale=0.24,clip]{pion_100001001100_vector1.eps}
\hspace{0.2cm}\includegraphics[scale=0.24,clip]{pion_100001001100_vector2.eps}
\hspace{0.2cm}\includegraphics[scale=0.24,clip]{pion_100001001100_vector3.eps}
\end{center}
\vspace*{-2mm}
\caption{Eigenvector components of the modified eigenvector problem for the
  pion ground state (l.h.s. plot), first
  and second excited states (center and r.h.s. plot). The plots correspond to
  the effective mass plateaus shown in figure 2.}
\end{figure}

\section{Conclusions and upcoming investigations}

In this article we have presented first results for pseudoscalar, scalar, vector
and pseudovector mesons with derivative quark sources and CI fermions on a quenched lattice. We employed
the variational technique with both Jacobi-smeared sources of a Gaussian type
as well as derivative-type sources.

It has been demonstrated that interpolators constructed with derivative quark
sources lead to an enhanced signal for a variety of different meson channels. For
pseudoscalar and vector mesons, the correlators constructed from such
interpolators display a significantly better overlap with excited states. In
both cases excited states could be observed that were not found with the
smaller basis.\\ For the scalar and pseudovector mesons, overlap with the ground states can be
improved which leads to ground state masses with significantly smaller
statistical errors.

We have clearly demonstrated that derivative quark sources provide an
essential enlargement of the basis of interpolators in excited meson
spectroscopy. Although we are convinced that derivative interpolators will be
an important tool also in the dynamical case, it is probably too early to claim
a one to one correspondence of the interpolator content.

We are currently extending our calculations to dynamical configurations
obtained in the BGR dynamical CI project. Furthermore, we investigate the
construction of baryon interpolators from derivative quark sources and the
effects of link smearing on our interpolators.

\acknowledgments
This work has been supported in part by the ``Fond zur F\"orderung der
wissenschaftlichen Forschung in \"Osterreich'' (FWF DK W1203-N08 and P19168-N16).

\bibliography{bibtex.bib}

\end{document}